# Zero-point energy of vacuum fluctuation as a candidate for dark energy versus a new conjecture of antigravity based on the modified Einstein field equation in general relativity

Guang-jiong Ni

Department of Physics, Fudan University, Shanghai, 200433, China

Department of Physics, Portland State University, Portland, OR97207, U.S.A.

Email address: pdx01018@pdx.edu

In order to clarify why the zero-point energy associated with the vacuum fluctuations cannot be a candidate for the dark energy in the universe, a comparison with the Casimir effect is analyzed in some detail. A principle of epistemology is stressed that it is meaningless to talk about an absolute (isolated) thing. A relative thing can only be observed when it is changing with respect to other things. Then a new conjecture of antigravity — the repulsive force between matter and antimatter derived from the modified Einstein field equation in general relativity — is proposed. This is due to the particle – antiparticle symmetry based on a new understanding about the essence of special relativity. Its possible consequences in the theory of cosmology are discussed briefly, including a new explanation for the accelerating universe and gamma-ray bursts.

## I. INTRODUCTION

The past decade has witnessed a remarkable progress of cosmology and astrophysics. Especially, the precise data provided by the Hubble Space Telescope (HST) key project, the Wilkinson Microwave Anisotropy Probe (WMAP) team, the Sloan Digital Sky Survey (SDSS) and the discovery of acceleration of the universe expansion, etc. have shaped an observable universe quite quantitatively. The present understanding can be summarized basically as follows (see, e.g. [1], [2]):

(a) The Hubble law — the linear relation between the redshift z of a galaxy and its distance D from earth

$$z = \frac{H_0}{c} D \quad (1)$$

has been verified up to D ~ 1000 Mpc with the Hubble constant:

$$H_0 = 72 \text{ km} / \text{s.Mpc} \quad (2)$$

(b) Since 1998, observations on the hundreds Type Ia supernovae with $z \geq 0.5$ reveal that the expansion of universe is accelerating rather than decelerating as speculated before.

(c) Defining a critical mass density of universe as

$$\rho_c = \frac{3H_0^2}{8\pi G} \approx 10^{-29} \, g/cm^3 \quad (3)$$

and a dimensionless parameter $\Omega_m = \rho / \rho_c$ with $\rho$ being the matter density of universe, cosmologists find that $\Omega_m \approx 0.3$. Moreover, the matter is composed of two parts, the visible baryon matter and the invisible dark matter. They are correlated in the space with their ratio being about 1: 6. The dark matter is still at large.

(d) The temperature fluctuations of cosmic microwave background radiation (CMB) (around the average value T = 2.725 K) measured accurately by WMAP show that the space-time of universe is flat, which means

$$O = O_m + \Omega_\Lambda = 1 \tag{4}$$

in the Friedmann model. Here the parameter

$$\Omega_\Lambda = \frac{c^2 \Lambda}{3H_0^2} \tag{5}$$

was introduced via a cosmological constant $\Lambda$ adding to the Einstein field equation of general relativity (GR) as follows[3]:

$$R_{\mu n} - \frac{1}{2} g_{\mu n} R - \Lambda g_{\mu n} = -8 p G T_{\mu n}. \tag{6}$$

(e) Analyses from WMAP, SDSS and the accelerating universe are focused on a value of $\Omega_\Lambda \approx 0.7$, which verifies the flatness of space as shown by Eq. (4) and gives an observational value of $\Lambda$:

$$\Lambda_{obs} \approx 1.3 \times 10^{-52} \text{ m}^{-2} \tag{7}$$

## II. DARK ENERGY AS THE ZERO-POINT ENERGY IN VACUUM

The term "dark energy" was coined in 1990 to reflect the strange feature of $\Lambda$, which makes a positive contribution to mass density as

$$r_? = \frac{c^2 \Lambda}{8 p G} \tag{8}$$

whereas it exerts a negative pressure as

$$p_? = -\frac{c^4 \Lambda}{8 p G} \tag{9}$$

In other words, its "equation of state" can be written as

$$p = w\, r\, c^2 \tag{10}$$

(the subscript $\Lambda$ is omitted) with w = -1. To explain this strange property, on the right-hand side of Eq. (6), a quantum average of the classical energy-momentum tensor $T_{\mu\nu}$ in the vacuum, i.e., the zero-point energy associated with the vacuum fluctuations (ZPEVF) $\langle T_{\mu\nu} \rangle_{vac} = -r_{vac} g_{\mu\nu}$, $\langle T_{00} \rangle_{vac} = r_{vac}$ is added to $T_{\mu\nu}$. One finds an effective equation as

$$R_{\mu\nu} - \frac{1}{2} g_{\mu\nu} R - \Lambda_{eff}\, g_{\mu\nu} = -8 p G T_{\mu\nu} \tag{11}$$

where

$$\Lambda_{eff} = \Lambda + 8 p G r_{vac} \tag{12}$$

should be identified with the $\Lambda_{obs}$ shown by Eq. (7).

There are two possible interpretations:

(1) Assume the classical cosmological constant $\Lambda = 0$. Then $\Lambda_{eff}$ is totally stemming from the ZPEVF:

$$\Lambda_{eff} = \Lambda_{obs} = 8 p G r_{vac} \tag{13}$$

According to quantum field theory (QFT), a kind of particle with mass m and spin j makes a contribution to the density of ZPEVF being [4, 5] ($\hbar = c = 1$):

$$\rho_{vac} = \frac{1}{2}(-1)^{2j}(2j+1)\int \frac{d^3k}{(2\pi)^3}\sqrt{k^2+m^2} \tag{14}$$

For the case of j = 0, m ≈ 0, the theoretical estimation reads:

$$\rho_{vac}^{theor} \approx \frac{1}{4\pi^2}\int_0^{k_{mac}} k^3 dk \approx \frac{1}{16\pi^2}(m_P c^2)^4 \approx 1.4 \times 10^{74} \, (GeV)^4 \tag{15}$$

Here, in order to control the ultraviolet divergence of integral, a cut-off $\hbar c k_{max} \approx m_P c^2 = \left(\frac{\hbar c}{G}\right)^{1/2} c^2 = 1.22 \times 10^{19} \, GeV$ (the Planck energy) is introduced.

On the other hand, combination of (13) and (7) yields:

$$\rho_{vac}^{obs} \approx 3 \times 10^{-47} \, (GeV)^4 \tag{16}$$

Then we see a ridiculous ratio being

$$\left(\rho_{vac}^{theor} / \rho_{vac}^{obs}\right) > 10^{120} \tag{17}$$

(2) Alternatively, noticing that the theoretical value of $\rho_{vac}$ can be negative as shown in (14) with j = ½, one may assume that the classical $\Lambda(\neq 0)$ and quantum $\rho_{vac}$ in (12) are cancelled each other so delicately that only a tiny value of $\Lambda_{eff}^{obs}$ is left as shown by (7). This fine tuning mechanism is called as the anthropic principle — we live where we can live.

### III. DARK ENERGY VERSUS CASIMIR EFFECT

The biggest mystery of dark energy lies in its strange property of negative pressure which implies some "antigravity" and leads to accelerating expansion of universe. More generally, the negative pressure is characterized by an equation of state (10) with the parameter w < -1/2 [6]. The model of cosmological constant (with w = -1) means the dark energy is uniform in space and constant in time.

Alternatively, the "quintessence" model with w > -1 means that the dark energy could be described by some dynamical scalar field which may vary with space and decay with time. Other models with w < -1 might be the sign of really exotic physics or some modification of GR. However, up to now, the data analyses still favor the simplest case of w = -1.

On the other hand, a number of physicists doubt that there is something wrong in the concept of dark energy (see, e.g. [7]). Moreover, some new theories are put forward, e.g. [8], trying to substitute the dark energy.

In this paper we wish to argue that the ZPEVF cannot be a candidate for dark energy. It was said that some experiment like the Casimir effect can be performed in laboratories, providing some evidence of ZPEVF to explain the dark energy. We don't think so. Let us have a look at the Casimir effect.

As is well known, the Casimir force $F_c$ is the gradient of Casimir energy $E_c$:

$$F_c = -\frac{\partial}{\partial d} E_c = -\frac{\pi^2 \hbar c L^2}{240 d^4} \tag{18}$$

And $E_c$ is the difference between ZPEVF of electromagnetic field in the presence of two (square) metal plates (with side length L and distance d between them) and that in the absence of them [9, 10]:

$$E_c = \sum_{\vec{k},a} \frac{1}{2}\hbar w_{\vec{k},a} \text{ (presence of plates)} - \sum_{\vec{k}',a'} \frac{1}{2}\hbar w_{\vec{k}',a'} \text{ (absence of plates)}$$

$$= \frac{1}{2}\hbar c L^2 \int \frac{d^2 k_{//}}{(2\pmb{p})^2} \left\{ |\vec{k}_{//}| + 2\sum_{n=1}^{\infty}\left(k_{//}^2 + \frac{n^2 \pmb{p}^2}{d^2}\right)^{1/2} - 2\int_0^{\infty} dn \left(k_{//}^2 + \frac{n^2 \pmb{p}^2}{d^2}\right)^{1/2} \right\} \quad (19)$$

Note that: (a) The wave vectors parallel to plates, $\vec{k}_{//}$, are the same for two cases (presence and absence of plates); (b) The wave numbers vertical to plates, $k_{\perp}$, are discrete (with n being discrete integer numbers) in the presence of plates whereas continuous (with n varying continuously) in the absence of plates; (c) For each fixed wave number, there are two modes of transversal polarization of field except for the case of $k_{\perp} = 0$ (n = 0) where only one mode (vertical to plates) exists. (d) The expression (19) is finite as shown by (18) and verified by experiments, see, e.g., [11].

Eq.(19) can be calculated either by introducing some cutoff function (e.g.[9]) or by resorting to a rigorous mathematical theorem — the Plana summation formula — as discussed in [12] (see also [10]). The reason why the theoretical result is free from any ambiguity is because $E_c$, being a difference, has neither ultraviolet divergence (when k → ∞) nor infrared divergence (when k → 0). In fact, the tiny contribution to $E_c$ is just coming from the difference of a sum and an integral with each of them being finite at k→0 region.

By contrast, aiming at an explanation for dark energy, the expression for the density of ZPEVF, Eq.(14), was written down in an ambiguous manner from the beginning. It is divergent at k→∞ limit. So, as a next step, an arbitrary cutoff was chosen to be the Planck energy (~ $1.22 \times 10^{19}$ GeV).

Third, actually, no one knows exactly how many varieties of different fields should be added?

Fourth, no one really understands why this positive $\pmb{r}_{vac}$ would lead to negative pressure of dark energy?

Fifth, no one can concretely explain how this negative (but isotropic) pressure exerts an outward repulsive force on a supernova to render it accelerating?

Last, but not least, why the theoretical estimation of $\pmb{r}_{vac}$ differs from that of observation so absurd as shown by (17)?

Physics is an accurate science as shown by the Casimir effect. The whole history of physics has been proving that a correct theory should be able to account for the experiments quantitatively. Once there is certain but small discrepancy between them, it must imply something of less importance being ignored in the theory. If the discrepancy turns to be large in quantity or even out of control, it means that something of importance must be wrong. Now we should regard Eq.(17) as a serious warning that we have been wrong in a fundamental way.

The reason why the evaluation for Casimir effect is correct whereas that for dark energy, Eq.(14), is wrong lies in the following clue: In the later case we were talking about an absolute energy of the cosmos space without a comparison between two configurations (as that in the Casimir effect). Thanks to researches on the dark energy over decade, we now have a very strong argument supporting the principle of relativity in physics (also in epistemology, see [13]) — No change, no information. Any information (e.g., the energy, the momentum or the position of a particle, etc) is created right at the occurrence of a changing process, i.e., at the measurement performed by the observer on the object.

In fact, numerous experiments, especially that aiming at clarifying the Einstein-Podolsky-Rosen paradox, all lead to the same conclusion that the concrete form of correlation in a quantum state exhibits itself as a direct consequence of measurement right at the destruction of original entanglement of state — there is no information existing before and after the measurement (for detail, see [10]).

Maybe many physicists still not believe in this principle mentioned above. Let us think about a simplest example: A particle with mass m is resting in the laboratory (S frame). I can say that it has no kinetic energy as long as I am staying in the S frame too. However, once I jump onto a train (S' frame) moving with a velocity v with respect to S, I will say that the particle has a kinetic energy $E = \frac{1}{2}mv^2$. The question is: Neither I nor anybody had done any work on the particle. Where its energy is coming from? This simple example shows that it is meaningless to talk about the absolute energy of an isolated object. Only when two particles are colliding each other, can the energy (and momentum) transfer be unambiguously calculated either in S or in S' frame. The energy (momentum) conservation law is meaningful exactly only right at the occurrence of collision, not before or after. Moreover, in fact, the definitions of energy and momentum are precisely meaningful only when they are conserved.

There is another problem in the theory of ZPEVF as dark energy — the theory of GR had been mixed with the QFT. We will discuss it after we further clarify the classical nature of GR in the next section.

## IV. ANTIMATTER AND A NEW MODEL FOR UNIVERSE EXPANSION

We wish to propose an alternative explanation for the accelerating universe, i.e., antigravity without invoking the dark energy. In our opinion, we should start from scratch. After studying the theory of special relativity (SR) for decades, we have been learning the essence of SR being nothing but the equal existence of particle and antiparticle [14, 15, 10, 13]. Many physicists thought that the so-called CP violation in weak interactions might trigger (in an early epoch of universe expansion) the asymmetry of particles and antiparticles — the former surpass the latter in number up to a factor like $10^8$-$10^9$. We don't think so. The CP inversion is equivalent to the so-called "time reversal", but the latter is essentially not a "time reversal" (which is merely a misnomer) but a "motion reversal" (see [16], [10]). Hence the tiny CP violation has nothing to do with the symmetry between particle and antiparticle. Rather, we believe that in the entire universe, the number of particles is equal to that of antiparticles.

Next, thanks again to SR combining with quantum mechanics (QM) and evolving to relativistic QM (RQM), we notice that there is a mass inversion ( m → -m ) symmetry in RQM to reflect the particle – antiparticle symmetry [17]. This symmetry, together with an interesting discussion of "negative mass paradox" raised in [18], leads to a natural generalization of Newton's gravitation law [17]:

$$F = \pm G \frac{m_1 m_2}{r^2} \quad (20)$$

Here the minus (plus) sign means the attractive (repulsive) force between matters or antimatters (matter and antimatter). Note that the mass of matter or antimatter is always positive. Eq.(20) can be deduced from the Einstein field equation in GR by modifying its right-hand side as [17]:

$$T_{\mu\nu} \to T_{\mu\nu}^{\text{eff}} = T_{\mu\nu} - T_{\mu\nu}^c \quad (21)$$

where $T_{\mu\nu}$ ($T_{\mu\nu}^c$) is the energy-momentum tensor of matter (antimatter) at the space point $x_m$. This modification makes Einstein equation invariant under the transformation m → - m.

In ref [17], some further conjectures are proposed as follows:

(a) The big bang created equal quantities of particles and antiparticles. The repulsive force between them triggered the inflationary expansion. But antiparticles had a head start and flew away faster than particles did. The latter lagged behind to some extent and our galaxies gradually evolved out of them. On the other hand, distant stellar objects might be evolved from antiparticles and some of them may be just those observed supernovae undergoing acceleration caused by the repulsive force exerted by inner galaxies composed mainly of matters.

(b) The flatness of our universe is critically depending on two factors: first, the inflationary expansion; second, the nearly equal densities of matter and antimatter in a large part of universe after the inflationary expansion. (See Appendix and Fig.1).

(c) In the intermediate region where matter and antimatter are overlapping, there may be some probability of collisions between matter and antimatter. Because of the long range repulsive force between them, these collisions may be of grazing type and so the annihilation process occurs at short distance would have some special character. Maybe it could explain the mechanism of gamma-ray- bursts (GRBs), the latter are distributed at remote distance on a cosmological scale and they tend to have a roughly constant explosive energy (especially for GRBs with typical lifetime T ~ 20s. Most likely, the gamma rays are emitted within two small-angle jets).

(d) Furthermore, it is observed that the star formation rate is rising steadily from remote distance with redshift z ~2 to z > 6 [19]. We guess this phenomenon is a reflection that the domination of antimatters is increasing steadily there. Since the coexistence of matter and antimatter would suppress the fluctuations inside the mixture of them and thus suppress the formation of stars, we guess there might be some anti correlation between the star formation rate and the spatial distribution of GRBs.

Fig.1 shows a schematic diagram of universe expansion. It is a separating process of matter from antimatter with (the cosmological) time. Hence their overlapping region decreases gradually and its two boundaries can be estimated via the measurements on the farthest and nearest GRBs (located at $G_f$ and $G_n$) respectively. Our Earth is located near the center of matter region because: (a). The distribution of GRBs is isotropic in all directions. (b). The CMB has a bipolar anisotropy showing the velocity of Earth being only V = 365 km/s with respect to the "cosmos rest frame" (CRF), whose origin should be the point surrounded by CMB isotropically as shown by two symmetric sources on the surface at $t_A = 3.8 \times 10^5$ y. Indeed, if resorting to the Hubble law, Eq.(1), and substituting the value of V = 365 km/s into the z = V/c, a distance of Earth from the origin of CRF can be estimated to be 5 Mpc which accounts only 0.2% of the length scale of matter region ( estimated to be 2000 Mpc at z =0.5). (c). Lucky enough, this location of Earth ensures a safer environment for human being's evolution. Other places in the universe would be suffered from more dangerous radiations emitted by nearer GRBs in the past billion years.

## V. SUMMARY AND DISCUSSION

(a) The essence of SR or that of the principle of relativity has two linked aspects, one in physics and one in epistemology [13]. The former is used to establish Eqs. (20) and (21) while the latter forms the main crux of our argument of why the ZPEVF cannot be the dark energy.

(b) The physical essence of SR is nothing but the symmetry of particle and antiparticle, which can be stated in one of two equivalent invariance, the invariance under the newly defined space-time inversion (x → -x, t → -t) at QM level [10] or that under the mass inversion (m → -m). The latter can be used either at QM level or at classical level. Being a classical field theory and treating only the gravitation among matters, the GR should be and can be modified to incorporate antimatter so that a new approach of big bang theory of cosmology might be established from scratch, treating matter and antimatter on an equal footing. However, even the modification shown in (21) has been made, it clearly refuses the addition of a quantum average of ZPEVF (even if it is considered ambiguously) to the classical energy momentum tensor because the contribution from matter will cancel that (if considered separately) from antimatter ( because both of them should have the same sign), a conclusion just in contradiction with the prediction of Eq.(14) by QFT (where the ZPEVF of particle and antiparticle is considered as a whole). So we see that the discussion of (11) through (17) is merely confusion in concept to combine a classical theory with quantum theory incorrectly. We still have no correct QFT for gravitation yet.

(c) Every law in physics is a discovery about the objective nature, also an innovation of human being as the subject. An isolated thing is absolute or abstract in the sense of being devoid of any cognition. We can only construct a meaningful theory based on (at least) two opposite things. Usually, in an experiment, one of them is just the apparatus. However, in QM, in order to describe an abstract quantum state vector $|y\rangle$ of a particle concretely, one has to resort to a basis vector $|x,t\rangle$ (which represents a "fictitious apparatus", see [10, 13]) for obtaining the wave function:

$$y(x,t) = \langle x,t|y\rangle \tag{22}$$

which is nothing but a probability amplitude of "fictitious measurement" to detect the "potential possibility" of finding the particle at position x and time t. The wave function is really a great innovation to endow the QM a prediction power before the real measurement is made. Another example is the Casimir effect. The existence of two plates allows us to compare the ZPEVF in two configurations (the presence and absence of two plates). In this case, every thing is finite and can be calculated unambiguously. By the way, all contributions of ZPEVF of massive particles to Casimir effect can be ignored.

(d) The divergence, i.e., the emergence of infinity has two aspects in mathematics: (1) it is a huge number; (2) it is uncertain. However, we physicists often emphasized the first one whereas overlooked the second one. We often introduced some cutoff to render the divergence under control like equation (15). Thus we overlooked the uncertainty in mathematics implying a warning in physics — it implies that we expected too much in trying to calculate something beyond our ability, or most likely, that thing doesn't exist at all. In short, we were wrong in some basic way.

(e) For example, in QFT, the divergence emerges every where. The situation was put by Dirac in the following words: (see [20])

"It would seem that we have followed as far as possible the path of logical development of the idea of QM as they are present understood. The difficulties, being of a profound character, can be removed only by some drastic change in the foundation of the theory, probably a change as drastic as the passage from Bohr's orbit theory to the present QM."

It seems to us that one important necessary change in basic concept is just the following answer to the famous Einstein's question: "Is the moon there when nobody looks?"[21] The "reality' should be defined at two levels: Originally, every thing contains no information at the level of "thing-in-itself" isolated from other things. Then it turns into the "thing-for-us" during (not before or after) a certain measurement by us, reflecting a series of observed phenomena. Every thing is infinite whereas our experimental data and theoretical interpretations must be fixed and finite before they can be meaningful. We had expected too much in QFT and been bothered by divergences for decades. Eventually, we got rid of them just because we began to distinguish what is existent thing whereas what is not. And we began to understand that we may get more only after we expect less. (See [13, 22, 23].)

(f) The particle-antiparticle symmetry prevails throughout the entire universe but it is spontaneously broken into two regions. We are living in the particle region and, very likely, near its center.

(g) Nowadays, one of the most important frontiers of physics lies in the fast developing field of cosmology and astrophysics. Among the greatest unsolved problems in science [24], the mystery of dark energy occupies a prominent position.

We are lucky to encounter a divergence like Eq.(14) once again which unveils a discrepancy between theory and observation as in (17) that is unprecedentedly huge. This should be regarded as a new, serious warning from Mother Nature. It's time for us to listen to her.

Acknowledgements: I am grateful to S. Q. Chen, P. T. Leung, E. J. Sanchez and Z.Q. Shi for helpful discussions.

APPENDIX. ACCELERATING UNIVERSE EXPLAINED BY MODIFIED EINSTEIN FIELD EQUATION

If similar to the Friedmann model [3], now the Einstein field equation ((6) without the $\Lambda$ term) with modification (21) is tentatively solved for the time (the cosmological time coordinate) evolution of a dimensionless cosmological scale factor R(t) in the Robertson-Walker metric. Being a measure of the deviation from the Hubble law, the time derivative of Hubble parameter $H(t) = \dot{R}(t)/R(t)$ could be derived approximately as (c =1):

$$\dot{H}(t) = \frac{d}{dt}H(t) = -4\pi G[(\boldsymbol{r} - \boldsymbol{r}^c) + (p - p^c)] \quad (A.1)$$

Here $\boldsymbol{r}(\boldsymbol{r}^c)$ and $p(p^c)$ are the mass density and pressure of matter (antimatter) in the flat cosmos with zero curvature. However, because antimatter flew away faster than matter, depending on the initial condition at the end of inflationary expansion era, there should be two co-moving coordinate systems related to them separately. This complication renders Eq. (A.1) and the following three possibilities only meaningful qualitatively:

(a) In the matter dominant region nearer to our galaxy, $\boldsymbol{r}^c \sim p^c \sim 0$, $\dot{H}(t) < 0$, the universe expansion is decelerating.

(b) In the intermediate region of universe where matter and antimatter are overlapping with nearly equal densities, $\boldsymbol{r} \sim \boldsymbol{r}^c$, $p \sim p^c$, $\dot{H}(t) \sim 0$, the linear relation of Hubble law remains valid and shows a coasting cosmos that is neither accelerating nor decelerating.

(c) In the outer region of universe where the antimatter dominates, $\boldsymbol{r} \sim p \sim 0$, $\dot{H}(t) > 0$, the universe expansion began to accelerate as shown by the supernovae (with z > 0.5) observed after 1998.

Of course, a more rigorous treatment is needed. But it seems likely that the cosmological constant (dark energy) is merely a false appearance of antimatter distributed at the remote distance.

REFERENCES
[1] Particle Data Group, Review of Particle Physics, Phys. Lett. B592(2004) 1-1109.
[2] M. Tegmark et al. Phys. Rev. D 69(10) 2004,10351.
[3] S. Weinberg, Gravitation and Cosmology (John Wiley, 1972).
[4] P. West, Introduction to Supersymmetry and Supergravity (World Scientific, Singapore, 1990).p.16.

[5] C. Beck and M.C. Mackey, ArXiv:astro-ph/0406504.
[6] R.R. Caldwell, Physics World, May, 2004, 37.
[7] S. Sarkar, Physics World, July, 2004, 15.
[8] E.W. Kolb, S. Matarrese, A. Notari and A. Riotto, ArXiv: hep-th/0503117.
[9] C. Itzukson and J-B Zuber, Quantum Field Theory (McGraw-Hill Inc. 1980) p.138-142.
[10] G-j Ni and S-q Chen, Advanced Quantum Mechanics (Press of Fudan University, 2000, 2003); English Edition was published by the Rinton Press, 2002.
[11] A. Lambrecht, Physics World, Sept. 2002, 29.
[12] G-j Ni, M. Zhang and X-w Gong, High Energy Physics and Nuclear Physics, 15(8) 695(1991); English Edition, 15(1991) 289.
[13] G-j Ni, Principle of Relativity in Physics and in Epistemology, accepted by "Relativity, Gravitation, Cosmology", ArXiv: physics/0407092.
[14] G-j Ni and S-q Chen, Journal of Fudan Univ. 35,325(1996) (in Chinese); and in "Photon and Poincare Group" (Edit: V.V. Dvoeglazov, NOVA Science Publisher, Inc. 1999) 145-169, ArXiv:hep-th/9508069.
[15] G-j Ni, Progress of Physics (Nanjing, China),23,484(2003) (in English).
[16] J. J. Sakurai, Modern Quantum Mechanics (Hohn Wiley & Sons,Inc. 1994).
[17] G-j Ni in "Relativity, Gravitation, Cosmology" (Edit: V.V. Dvoeglazov and A.A. Espinoza Garrido, NOVA Science Publishers, Inc.) Chapt.10, 123-136, ArXiv: physics/0308038.
[18] C. Will in "The New Physics" (Edit: P. Davies, Cambridge Univ. Press,1989) p.31.
[19] B.E. Schaefer, The Astrophysical Journal, 583:L67-L70(2003).
[20] J. J. Sakurai, Advanced Quantum Mechanics, (Addison-Wesley Publishing Company, Inc. 1967) P.295.
[21] N. D. Mermin, Physics Today, April, 38 (1985).
[22] G-j Ni, S-y Lou, W-f Lu and J-f Yang, Science in China (Series A) 41, 1206 (1998); ArXiv:hep-ph/9801264.
[23] G-j Ni, G-h Yang, R-t Fu and H-b Wang, Int. J. Mod. Phys. A16, 2873 (2001).
[24] J. R. Vacca, The World's 20 Greatest Unsolved Problems, Prentice Hall, 2005.
[25] J. Hjorth, C. Kouveliotou and S. Woosley, Physics World, Oct. 2004, 35.

Figure caption for Fig.1:

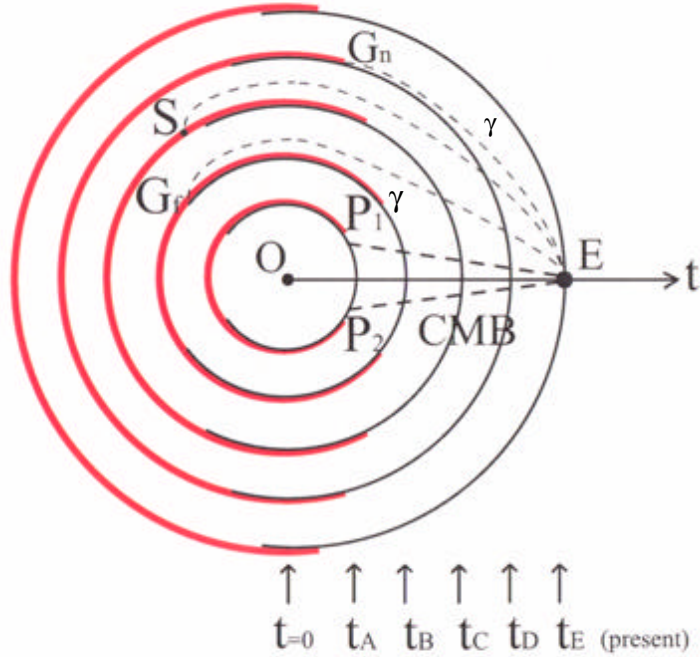

Schematic diagram (not in scale) of matter-antimatter universe evolving with the cosmological time t. The origin is the big bang at t = 0. The three-dimensional space of cosmos at t = const is simplified into a two-dimensional spherical surface, whose cross-section with this paper is shown as a circle with its right and left parts being filled by matter and antimatter separately. Their overlapping region is shrinking with time. Two points at the boundaries $G_f$ and $G_n$ denote the farthest and nearest GRBs observed on Earth (E) today ($t_E = 1.37 \times 10^{10}$ y). They can be tentatively identified with the GRB000131 (at distance D= $12.3 \times 10^9$ ly) and GRB030329 (D= $2 \times 10^9$ ly) respectively [25]. So they exploded at time $t_B = 1.4 \times 10^9$ y and $t_D = 11.7 \times 10^9$ y respectively. Similarly, the point S denotes a supernova with D ~ 7 x $10^9$ ly located in the antimatter region and exploded at time $t_C \sim 6.7 \times 10^9$ y. It was accelerating away from us as observed. Points $P_1$ and $P_2$ show two sources of CMB emitted at $t_A = 3.8 \times 10^5$ y.